# Precise control on morphology of ultrafine LiMn$_2$O$_4$ nanorods as supercapacitor electrode via two-step hydrothermal method

Niraj Kumar[a*], K. Guru Prasad[bc], T. Maiyalagan[b] and Arijit Sen[bc*]

[a] Kalasalingam Academy of Research & Education, Krishnankoil, 626126, India
[b] SRM Research Institute, SRM Institute of Science & Technology, Kattankulathur 603203, India
[c] Department of Physics & Nanotechnology, SRM Institute of Science & Technology, Kattankulathur 603203, India

[*] Corresponding Author E-mail: nirajunisci2k@gmail.com, arijit.s@res.srmuniv.ac.in

**Abstract**

We report three different synthesis routes while maintaining similar reaction conditions to choose an effective way to precisely control the growth of ultrafine one-dimensional (1D) LiMn$_2$O$_4$ in the form of nanorods. We developed a novel method of mixing the precursors through hydrothermal, yielding low dimensional precursors for effective solid state reaction to synthesize the nanorods. However, to achieve these, highly uniform $\beta$-MnO$_2$ nanorods were initially grown as one of the main precursors. The uniformity observed in as grown $\beta$-MnO$_2$ nanorods using hydrothermal technique help to attract minute LiOH particles upon mixing over its highly confined nano-regime surface. This facilitated the solid state reaction between MnO$_2$ and LiOH to develop one of the finest LiMn$_2$O$_4$ nanorods with diameters of 10-80 nm possessing high surface area of 88.294 m$^2$/g. We find superior charge storage behaviour for these finely ordered 1D nanostructures as supercapacitor electrodes in KOH with K$_3$Fe(CN)$_6$ as electrolyte in contrast to Li$_2$SO$_4$. A high pseudo-capacitance of 653.5 F/g at 15 A/g is observed using galvanostatic discharge time with high retention capacity of 93% after 4000 cycles. The enhanced charge storage property may arise from the redox couple [Fe(CN)$_6$]$^{3-}$/[Fe(CN)$_6$]$^{4-}$ and K$^+$ ions of the electrolyte. To the best of our knowledge, we demonstrate for the first time the effectiveness of a two-step hydrothermal method in tuning the supercapacitive behaviour of 1D LiMn$_2$O$_4$ in redox additive electrolyte.

## 1. Introduction

To meet the increasing energy demands, natural resources (*e.g.* fossil fuels) are being depleted at a faster rate whence is the need for an alternate energy resource [1-7]. Supercapacitors can be a promising alternative owing to its high energy density, stability, cyclability and power efficiency [8-11]. They are being used extensively in hybrid vehicles, as power stabilizers in electronic devices and renewable energy systems [12-15]. They can also provide high power when coupled with batteries and fuel cells [7, 16, 17]. Unfortunately, they suffer from low charge capacity, compared to Li-ion batteries [18]. Pseudocapacitors can be a solution to this which collectively enjoys the advantages of both supercapacitor as well as Li-ion battery [19-26]. Different materials like Li$_4$Ti$_5$O$_{12}$, LiMn$_2$O$_4$, Na$_x$MnO$_2$ etc. have been studied in recent time as cathode materials for asymmetric supercapacitor to increase their efficiency [27-33]. Among them, spinel LiMn$_2$O$_4$ (LMO) is one of the most promising cathode materials due to its environmentally benignity, thermal stability and low cost [34].

Preparation of LMO with high rate capability and capacitance is a challenge for researchers. Developing nanoscale materials is a solution to this and so the syntheses of nanorods, nanowires, nanotubes and nanochains have gained much interest among researchers [34-39]. The intercalation and de-intercalation of charges improves with the use of nanoscale materials which ensures good ionic and electronic conductivity [40]. Fehse et al. [41] have demonstrated ultrafast dischargeable $LiMn_2O_4$ thin film electrodes with pseudocapacitive properties for microbatteries. Benjamin et al. [42] have studied mesoporous $Li_xMn_2O_4$ thin film cathodes as lithium-ion pseudocapacitors. However, most of the electrolytes ($LiPF_6$ and $LiClO_4$) used for energy storage need inert atmosphere requiring high cost. Aqueous electrolyte is an alternative as it can be used in ambient condition. $LiMn_2O_4$ nanomaterials are being studied in aqueous electrolytes as promising electrode material [36-39]. Wang et al. [43] have shown spinel $LiMn_2O_4$ nano-hybrid comprising nanotubes, nanoparticles and nanorods as high-capacitance positive electrode material for supercapacitors in aqueous electrolyte. It may be noted that there are still very few reports on pseudocapacitive performance of $LiMn_2O_4$ nanorods in aqueous electrolyte. Moreover, there is a lack of proper synthesis technique to develop high quality $LiMn_2O_4$ nanorods for its commercialization.

In the current work, we followed a solid state reaction to grow $LiMn_2O_4$ nanorods of 10-80 nm diameters, after a novel hydrothermal route for effectively mixing the precursors utilizing as synthesized $\beta$-$MnO_2$ nanorods. Hydrothermal technique was preferred for its simplicity and effectiveness [44-47]. Hypothetically, this could be a first report on utilization of two-step hydrothermal process to achieve one of the finest 1D structures of $LiMn_2O_4$. Surface analysis through BET method and morphological analysis through FESEM were followed in selection of optimized synthesis steps from three different synthesis procedures at similar reaction conditions. The supercapacitive performance of the as synthesized $LiMn_2O_4$ nanorods was evaluated in two different aqueous electrolytes, namely $Li_2SO_4$ and KOH with $K_3Fe(CN)_6$ as redox additive. In the later case, we observed a much superior charge storage properties. For the first time, we demonstrate the promising pseudocapacitive property of $LiMn_2O_4$ nanorods in redox additive.

## 2.1 Chemicals

Potassium permanganate ($KMnO_4$), Sodium nitrite ($NaNO_2$), Sulphuric acid ($H_2SO_4$), Potassium hydroxide (KOH), Potassium ferricyanide ($K_3Fe(CN)_6$), Lithium sulphate ($Li_2SO_4$), Polyvinylidene fluoride (PVDF) and Super P carbon were procured from Sigma Aldrich and all the solutions were made with de-ionized water.

## 2.2 Growth of $\beta$-$MnO_2$ nanorods

In usual synthesis, 5 mM and 7.5 mM of $KMnO_4$ and $NaNO_2$, respectively were mixed in 2:3 molar ratios in 37.5 ml of de-ionized water and magnetically stirred for nearly one hour to obtain dark red coloured solution. Then, 2.5 ml solution of 0.4 M $H_2SO_4$ was prepared and added drop-wise at specified intervals of time so that there is a slow transition of this dark red coloured solution into light red. The solution was bolted inside 50 ml autoclave and heated at 170°C for 12 h. The autoclave was left to get cooled, normally. The final product was seen to be settled beneath the residual liquid. Finally, a reddish brown coloured product was collected after washing with de-ionized water for a couple of times and drying in air at 80°C. It was then heated for 350°C for 6 h to remove un-reacted impurities in way to obtain highly pure $\beta$-$MnO_2$ nanorods.

## 2.3 Growth of LiMn$_2$O$_4$ nanorods

For the synthesis of LiMn$_2$O$_4$ nanorods, 0.4868 g of as synthesized β-MnO$_2$ nanorods was thoroughly mixed with 0.12 g LiOH•H$_2$O in 40 ml of de-ionized water under continuous stirring. After half an hour, the solution was transferred to 50 ml teflon and sealed inside autoclave for hydrothermal process at 170°C for 12 h. Same as before, the final product was washed and dried in air to obtain fine black coloured powder. It was then kept for solid state reaction by heating at 750°C for 24 h. The final product was centrifuged, washed for several times and named as S1.

      In way to optimize the synthesis procedures, 4 mg LiOH•H$_2$O was added at the start of the β-MnO$_2$ nanorods synthesis process along with KMnO$_4$ and NaNO$_2$ and the same synthesis steps were followed. The as obtained product was then allowed for solid state reaction at 750°C for 24 h and named as S2 after its centrifugation and cleaning.

      In another typical synthesis, different synthesis steps were as followed. 0.4868 g of as synthesized β-MnO$_2$ nanorods was thoroughly mixed with 0.12 g LiOH•H$_2$O through vigorous grinding in agate mortar. Few drops of ethanol were added during grinding to make solution paste. The grinding was done until the mixture fully dried up at room temperature. The mixture was then kept in alumina boat and allowed for solid state reaction for 24 h at 750°C. The final product was centrifuged and washed for several times and named as S3.

## 2.4 Preparation of positive electrode for supercapacitor

For the preparation of cathode material 8:1:1 weight ratio of LiMn$_2$O$_4$, super P carbon and PVDF were mixed by grinding in agate. Super P carbon and PVDF were added respectively to improve the conductivity and adhesiveness of LiMn$_2$O$_4$ as the active material. N-Methyl-2-pyrrolidone was added into the mixture to form gelatinous solution. 304 grade steel plate with 0.3 mm thickness was used as substrate upon which 0.7 mg of as prepared electrode material was coated uniformly within area of 1 cm$^2$. The coated plate was then dried at 80°C for 12h.

## 2.5 Characterization

Structural properties were evaluated using XRD and Raman analyses through 'PAN analytical X' Pert Pro diffractometer and 'LASER Raman spectrometer', respectively. Physical overviews of the samples were made through FESEM and HRTEM analyses using 'Quanta 200 FEG FE-SEM', and 'HR-TEM, JEM-2010', respectively. Electrochemical studies were carried on Biologic SP300. A three electrode system was employed with silver/silver chloride (Ag/AgCl, 3M KCl) and platinum wire as reference and counter electrodes respectively for as prepared cathode material as working electrode.

## 3 Results and discussions

### 3.1 Structural and morphological analyses

To confirm the structure of the sample, obtained after the hydrothermal treatment of KMnO$_4$, NaNO$_2$ and H$_2$SO$_4$ compounds, XRD spectrum was obtained as presented in fig. 1a. The major diffraction peaks at 2θ = 28.6, 37.3, 41.1, 42.8, 56.6, 59.3, 67.2 and 72.3 are seen to be aligned in sequential order of intensity ascribing the planes (110), (101), (200), (111), (211), (220), (310) and (301), respectively with no other impurity peaks. This reveals highly pure tetragonal beta phase MnO$_2$ according to JCPDS 24-0735 [44]. Raman spectroscopy was followed to further strengthen the structural identification of as prepared β-MnO$_2$ sample. The

Raman spectra shown in fig. 1b, reads two prominent peaks at 574 and 644 cm$^{-1}$ due to vibrating and stretching modes of Mn-O bonding in MnO$_6$ octahedra structure confirming the tunnel structure of β-MnO$_2$ [48, 49]. The as synthesized β-MnO$_2$ are in nanorods shape with diameters in range of 10-40 nm as seen clearly from FESEM images presented in fig. 1c-d. The morphologies seen are strictly 1D and no other shapes are visualized which reveals ultrafine nature of the as synthesized β-MnO$_2$ nanorods.

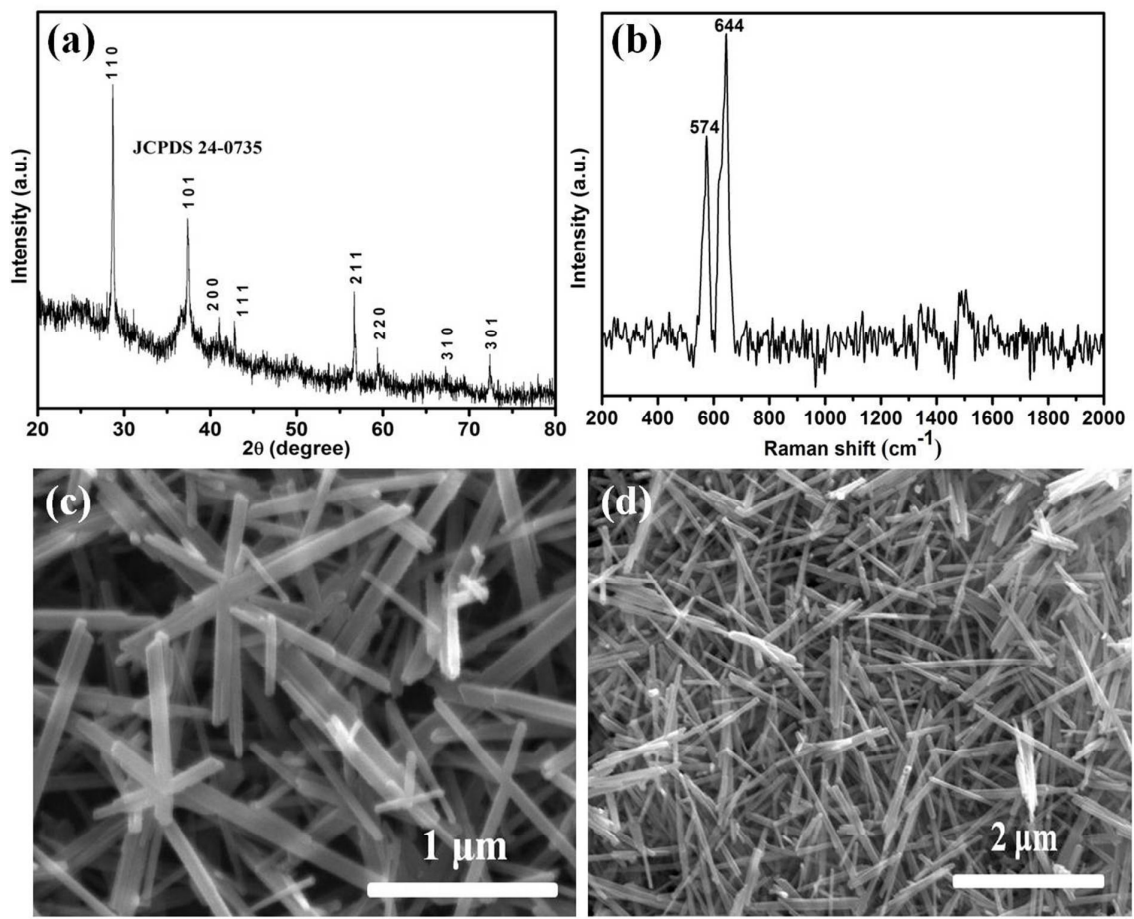

**Figure 1.** (a) XRD pattern (b) Raman spectra and (c-d) FESEM images of as prepared β-MnO$_2$.

Figure 2a, displays the XRD patterns for samples S1, S2 and S3. All the samples commonly exhibit eight major diffraction peaks at 2θ = 18.688, 36.077, 37.935, 44.01, 48.233, 58.194, 63.9 and 67.311 corresponding to planes (111), (311), (222), (400), (331), (511), (440) and (531) respectively. This strongly attributes to the cubic structure of LiMn$_2$O$_4$ according to JCPDS 88-1026. It may be noted that a constant phase for the different samples is observed in-spite of their different synthesis procedures. This indicates that a perfect optimization was achieved in all the three synthesis procedures due to the utilization of comparable amount of chemical precursors followed by the solid state reaction. In agreement with the XRD analysis, similar Raman spectra can be seen in fig. 2b for all the samples (S1-S3) in the spectral region of 100-1000 cm$^{-1}$. Prominent band at 640 cm$^{-1}$ in conjugation with two less intense bands at 611 and 569 cm$^{-1}$ is conspicuous for all the samples. The high intensity band at 640 cm$^{-1}$ is in close proximity with the band observed in Raman spectrum of β-MnO$_2$ shown in fig. 1b, which strongly indicates stretching modes of Mn-O bonding.

Octahedral MnO$_6$ unit of as prepared LiMn$_2$O$_4$ experiences vibrations in oxygen atoms from the spinel oxide ascribing to bands at 640 and 611cm$^{-1}$ [50, 51]. The vibrations relate to bond stretching between manganese and oxygen atoms (Mn-O) ascribing to O$_h^7$ spectroscopic symmetry incorporating $A_{1g}$ species. The band at 569 cm$^{-1}$ is seen to be not fully separated from the main intensity peak as observed for all the samples and appears like shoulder to the main high intensity band. This characteristic arises from the Mn$^{+4}$–O vibrations and resembles Lithium stoichiometry in LiMn$_2$O$_4$ [51]. The morphology of as prepared LiMn$_2$O$_4$ is revealed to be ultrafine nanorods of 10-80 nm diameters from the HRTEM images as displayed in figure 2c. In corollary to the XRD analysis, $d$-spacing of 0.48 nm in (111) plane is observed from the uniform fringes shown in HRTEM image (inset fig. 2c). The selected area electron diffraction (SAED) pattern in fig. 2d of single nanorod shows the miller indices (311), (400), (511) and (531) respective to the $d$-spacings matching with JCPDS 88-1026, further confirms for the formation of LiMn$_2$O$_4$.

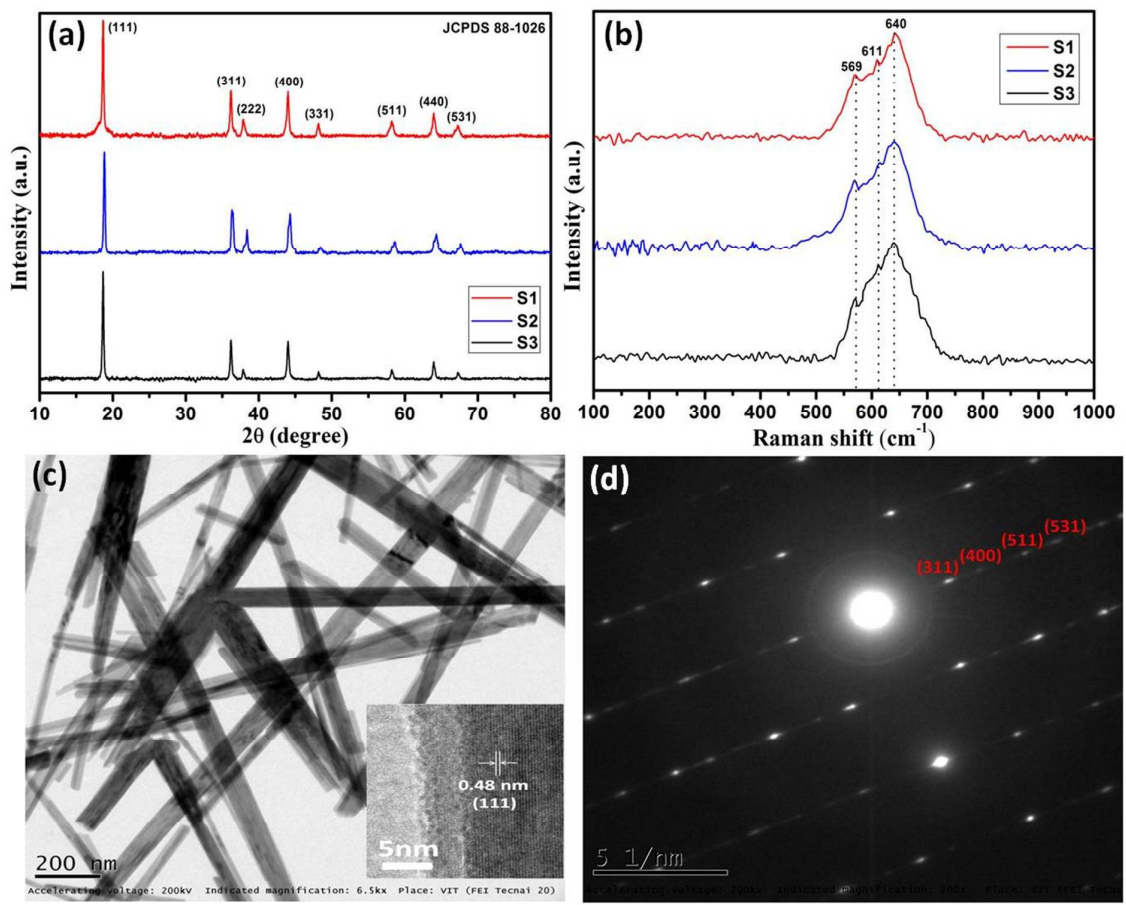

**Figure 2.** (a) XRD pattern and (b) Raman spectra of LiMn$_2$O$_4$ samples S1, S2 and S3; (c) HRTEM image with (inset in c) $d$-spacing of 0.48 nm in growth direction (111) and (d) SAED pattern of sample S1.

To explore a perfect synthesis route for the ultrafine nanorods growth, three different syntheses procedures were studied through FESEM analysis. Samples S3, S2 and S1 are the outcomes of the three syntheses procedures as mentioned above. From FESEM images of S3 (fig. 3a-b), it is clear that some microstructures are also present in conjugation with the nanorods. It is assumed that for proper feasibility of the solid state reaction at 750°C, the precursor (LiOH) must be interacted or adsorbed completely on to the surface of these as grown ultrafine β-MnO₂ nanorods. Unfortunately, this can not be achieved through simple uneven mixing in agate mortar as described earlier. Therefore, in our next attempt, we directly added the precursors at start of the hydrothermal reaction, so that while these nanorods grow, they could easily incorporate LiOH particles under the hydrothermal pressure. Fortunately, comparatively less or smaller microstructures are seen in FESEM images of sample S2 (fig. 3c-d). And finally, we followed two step hydrothermal processes, one to grow the β-MnO₂ nanorods and the second to homogenously mix the LiOH particles on to the surface of these nanorods. This resulted in the formation of ultrafine LiMn₂O₄ nanorods as seen in FESEM images of sample S1 (fig. 3e-f). Thus hydrothermal method is quite effective to obtain morphologies in nanoscale regime. The reason behind this could be the as developed hydrothermal pressure which forces the different precursors to interact with each other in way to release this pressure, thereby facilitating the chemical kinetics. To highlight the optimization process, a schematic has been presented in fig. 4 to account for the synthesis steps in preparation of samples S1, S2 and S3.

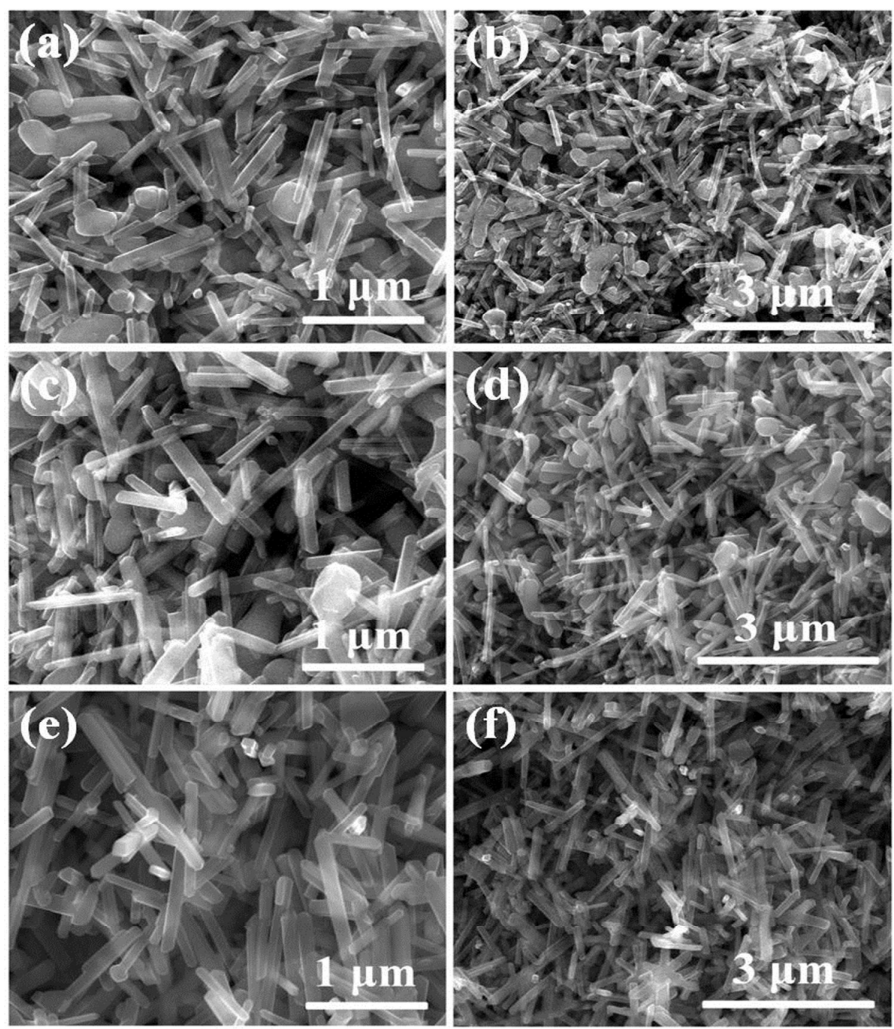

**Figure 3.** FESEM images of as prepared sample (a-b) S3, (c-d) S2 and (e-f) S1.

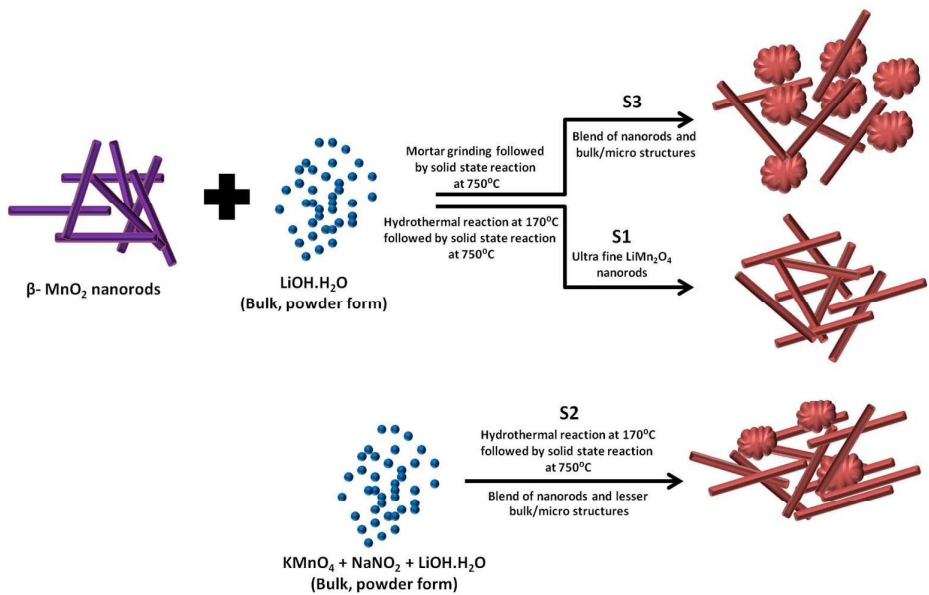

**Figure 4.** Schematic for the growth conditions of samples S1, S2 and S3.

A physical overview on the growth mechanism of ultrafine $LiMn_2O_4$ nanorods (S1) can be postulated based on the FESEM images shown in fig. 5(a-d) captured at times of 2, 6, 12 and 24 h, respectively, after the heat treatment at 750°C of LiOH and $β$-$MnO_2$ nanorods. It is clear that when the heating continued for 2 h then cloud like morphologies of the reagents is seen covering the 1D structures. On further increasing the time of heat treatment to 6 h and 12 h, the cloud like morphologies begins to disappear suggesting the diffusion of elementary particles into the interiors of 1D morphologies. This phenomenon predicts for the effective solid state reaction between LiOH and $β$-$MnO_2$ nanorods [52]. The chemical action at the boundaries of 1D structures increases with increasing time of heat treatment. Finally after 24 h of heat treatment, complete disappearance of cloud like morphologies is perceptible. This could be due to the fusion of elementary Li atoms into $MnO_2$ structure via solid state reaction giving rise to ultrafine $LiMn_2O_4$ nanorods.

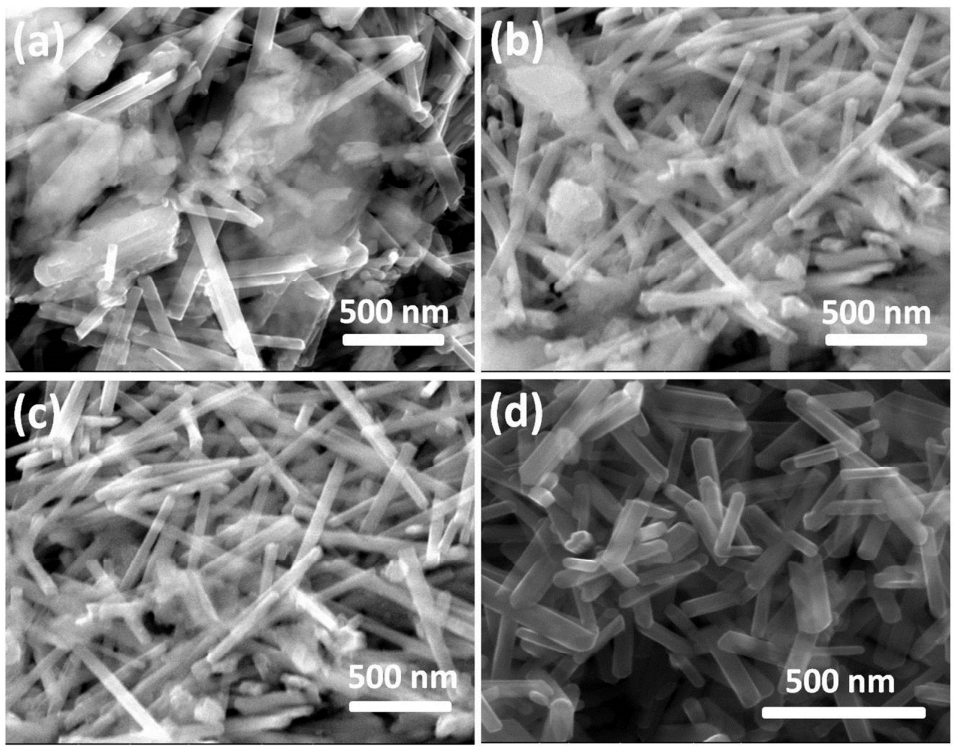

**Figure 5:** FESEM images of sample S1 after the solid state reaction between LiOH and β-MnO$_2$ nanorods at (a) 2 h, (b) 6 h, (c) 12h and (d) 24 h.

### 3.2 Surface analysis

Figure 6a-c shows the nitrogen (N$_2$) adsorption-desorption isotherms of the samples S1, S2 and S3, respectively. A gradual increment in volume adsorption is perceptible for all the samples. The adsorption-desorption curves coincide and resembles Type II for sample S1 (fig. 6a) and Type III for samples S2 and S3 (fig. 6b-c) [53]. The absence of adsorption-desorption hysteresis can be attributed to the multilayer formation similar to macroporous solids. Sample S1 may contain monolayer relating to Type II whilst for samples S2 and S3; convex isotherm is viable depicting Type III. These diversified isotherms can be attributed to the different morphologies as synthesized. The bending of isotherm occurs due to stronger interactions between adsorbates then compared to their interaction with the adsorbent surface. Therefore, for sample S2 and S3, formation of more multilayers composed of adsorbates is expected. It is notable that these samples are blend of micro (or bulk) and nanostructures. On the other hand, the ultrafine nanorods like structure (S1) exhibits almost linear isotherm due to the presence of stronger interaction between adsorbent and adsorbate giving the possibilities of only single layer formation. The volume adsorption for samples S1, S2 and S3 follows a decreasing order. In corollary, these fine nanorods (S1) shows high surface area of 88.294 m$^2$/g compared to the surface areas of 42.988 and 21.477 m$^2$/g for S2 and S3, respectively (table 1). This can be ascribed to the confined 1D morphologies in sample S1. Moreover, for sample S3, a little separation in absorption-desorption isotherm is conspicuous, which corresponds to stronger interaction between adsorbent and adsorbate than sample S2. This analysis confirms for larger micro or bulk structures presence in S3 than S2 and the most confined structures in S1 which is evident from the above morphological analysis. The BJH pore size distributions (PSDs) for samples S1, S2 and S3 are shown in fig. 6d-f, respectively. The PSD histograms mainly converge around 1-3 nm. The average pore diameters and pore

volume for sample S1, S2 and S3 are presented in table 1. Considerably small pore diameters can be attributed to the highly confined morphologies in nanoscale regime.

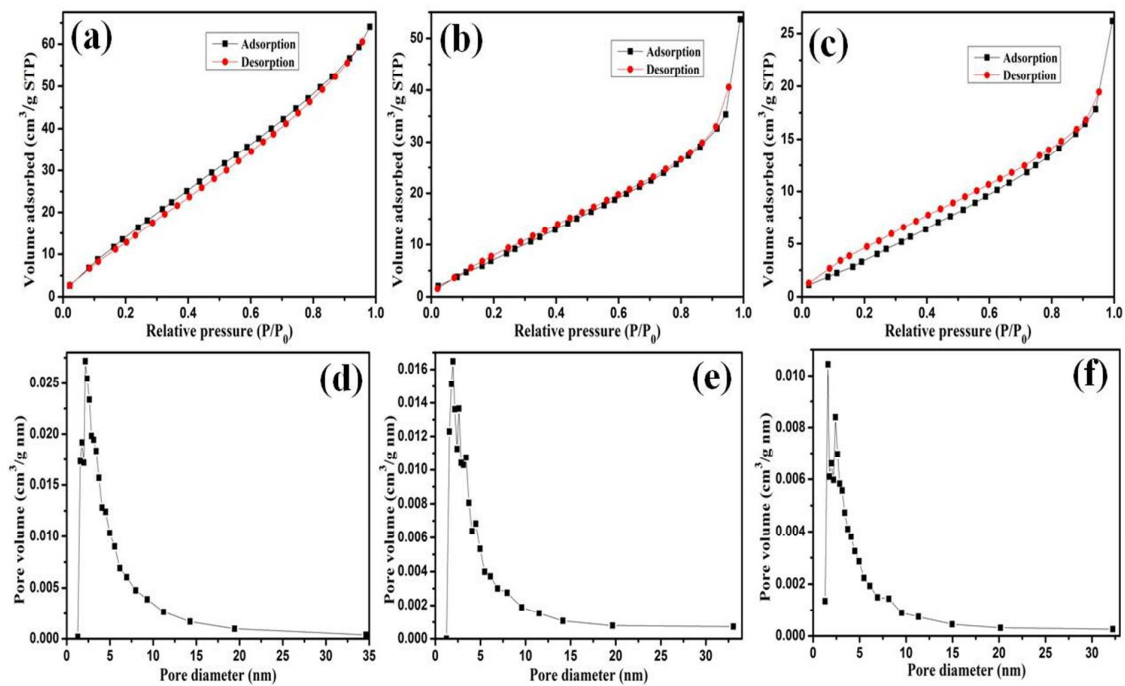

**Figure 6.** (a–c) Nitrogen ($N_2$) adsorption–desorption isotherms and (d-f) BJH pore size distributions (PSDs) of samples S1, S2 and S3, respectively.

**Table 1.** BET surface area, average pore diameter and pore volume of sample S1, S2 and S3.

| Sample | Surface area ($m^2/g$) | Average pore diameter (nm) | Average pore volume (cc/g) |
|---|---|---|---|
| S1 | 88.294 | 2.145 | 0.119 |
| S2 | 42.988 | 1.949 | 0.078 |
| S3 | 21.477 | 1.593 | 0.038 |

**3.3 Electrochemical studies**

**3.3.1 In $Li_2SO_4$ aqueous electrolyte**

A three electrode system comprising Ag/AgCl (3M KCl), platinum wire and as prepared electrode acting as reference, counter and working electrodes respectively were fabricated for electrochemical studies. The test was carried out in 0.5 M $Li_2SO_4$ aqueous electrolyte. Similar cyclic voltammograms are observed (fig. 7a) for all the samples S1-S3 at 5 mV/s in potential window of 0.2 to 1.2 V. This suggests for complete similarity in phase and structure of as prepared samples and agrees to the XRD and raman analyses. Figure 7b shows the cyclic voltammetry (CV) measurements in potential window of 0.2 to 1.2 V at different scan rates of 4, 6 and 8 mV/s. A pair of anodic oxidation peaks is observed at 0.88 and 0.98 V at positive current whereas a pair of cathodic reduction peaks is present at 0.80 and 0.90 V at negative current. Close symmetry of these two redox peaks pairs reveals for excellent reversibility of Li$^+$ extraction and insertion into as spinel LiMn$_2$O$_4$ nanorods. This may due for the ultra finesse morphology of nanorods favouring short diffusion path between electrode/electrolyte interfaces. The two anodic and cathodic reverse activities can be ascribed to two partitioned process of lithiation/de-lithiation as mentioned below [42]:

$$LiMn_2O_4 \leftrightarrow Li_{0.5}Mn_2O_4 + 0.5Li^+ + 0.5\ e^- \qquad (1)$$

The above process repeats twice to give final lithiation/de-lithiation process as:

$$LiMn_2O_4 \leftrightarrow Mn_2O_4 + Li^+ + e^- \qquad (2)$$

Furthermore, it can be visualized that there is close proximity of peaks in each pair as compared to sharp distinct peaks in the literature [18, 41-43]. Therefore two distinct peaks for each anodic and cathodic can be can be viewed to be one as they nearly resembles to only two capacitve redox peaks described in the literature [54]. The effect is more viable as seen in the CVs at higher scan rates of 16 and 32 mV/s (fig. 7b). This gives an indication for fast cyclic ability of energy system suggesting for ultrafast lithiation and de-lithiation process, which is a prerequisite for a superior capacitive performance.

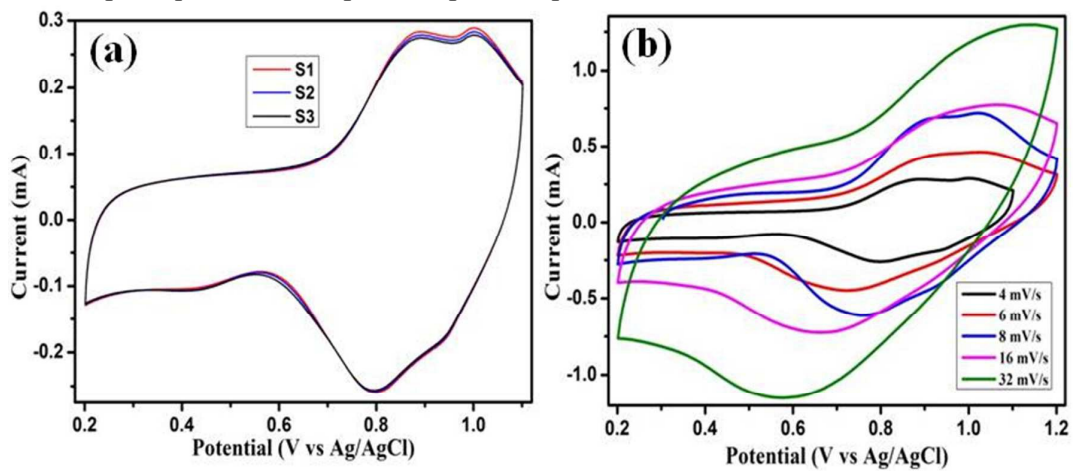

**Figure 7.** Cyclic voltammetry measurements for sample (a) S1, S2 and S3 at 5 mV/s and (b) S1at different scan rates in 0.5M Li$_2$SO$_4$ electrolyte.

When the scan rate for CV was increased, the area under CV curve also gets increased, which can be attributed for promising capacitive performance of as prepared LiMn$_2$O$_4$ nanorods. However, the discharge current tends to vary during CV, so it's unfair to predict capacitance using the CV curve. Specific capacitances C (F/g) for samples S1-S3 were calculated using galvanostatic charge/discharge measurements carried out in potential range from 0.2 to 1.2 V according to [43]:

$$C = I\ \Delta t\ /\ m\ \Delta V \qquad (3)$$

Here, current (I) is in Ampere, discharge time ($\Delta t$) is in seconds, mass of the active material (m) is in grams and applied potential range ($\Delta V$) is in volts. Initial galvanostatic charge/discharge measurements for sample S1 at 0.5, 1.0, 1.5 and 2 A/g is shown in fig. 8a. Specific capacitance (C), were calculated to be 144.5, 119.7, 111.5 and 98.7 F/g at current densities of 0.5, 1.0, 1.5 and 2 A/g for sample S1. High-capacitance value observed can be attributed to the large amount of charge accumulation at electrode/electrolyte interfaces during lithiation/de-lithiation process as described above. Capacity retention after 2000 cycles was estimated for all the samples as displayed in fig. 8b. Capacity fading was observed with increasing current density, which can be a usual characteristic of supercapacitors. Highly stabilized capacitive performance has been encountered with a high retention capacity of

97.3% after 2000 cycles for sample S1. This galvanostatically calculated capacitance using discharge time could possibly be one of the very few reports and suggests the material to be promising cathode for asymmetric supercapacitor [18, 41-43]. For sample S2 and S3, 86% and 81% of capacity retention is observed, respectively (fig. 8b). In spite of possessing similar phase and structure, these materials (S1-S3) vary in their prolonged cycling behaviour. Morphological variation could be the viable reason. Sample S1 performed the best followed by S2 and S3. Sample S1 possessed more confined and uniform morphologies which helped it to exhibit very high surface area as discussed above followed by S2 and S3. As a result sample S1 as electrode was able to interact well with the electrolyte when used through shortening of diffusion length and enhanced surface activity. The study clearly indicates that highly confined nanoscale structures are well suited for charge storage. First 20 cycles of galvanostatic charge-discharge profiles for sample S1 is presented in the inset fig. 8b to show a promising stability at initial cycling processes.

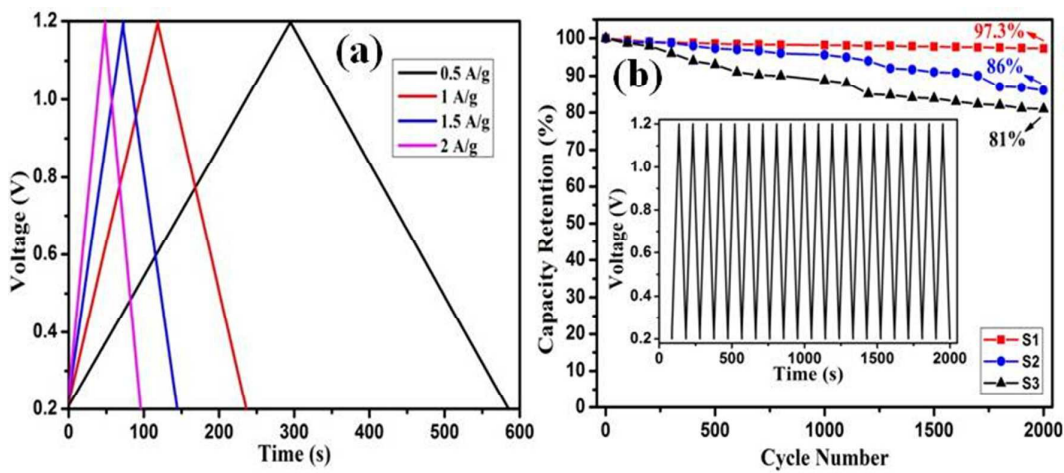

**Figure 8.** (a) Galvanostatic charge/discharge cycles at different current densities for S1 and (b) Capacity retention versus cycle number at 2 A/g for S1, S2 and S3 with inset (b) first 20 cycles of Galvanostatic charge/discharge for S1 in 0.5M $Li_2SO_4$ electrolyte.

### 3.2.2 In KOH aqueous electrolyte with $K_3Fe(CN)_6$ as redox additive

In the above electrochemical analysis, two major drawbacks persist. First, being the low capacity and the other being a very high discharge time due to its low current density. A supercapacitor is known to deliver promising energy at relatively small time compared to batteries. To achieve this, high charge storage property is expected at promisingly high current density [55]. In account of this, we followed the same electrochemical analysis (the three electrode system) with only change in the electrolyte under different potential window. In advancement of our previous report [55], we used 3 M KOH with redox additive as 0.1 M $K_3Fe(CN)_6$ to test the charge storage properties of as synthesized $LiMn_2O_4$ nanorods. Like the previous case, here also close similarity is observed (fig. 9a) in CV patterns for all the samples at 15mV/s in the potential window of -0.4 to 0.6V. This again confirms the close proximity in phase and structure of as prepared samples. The cyclic voltammograms at different scan rates of 5, 10, 15, 20, 25 and 30 mV/s for sample S1 are shown in fig. 9b. The low scan rates are incorporated to account for the exact positions of redox peaks, which normally disappear at higher scan rate. In the potential window of -0.4 to 0.6V, oxidation (anodic) and reduction (cathodic) peaks can be seen approximately at 0.52 and 0.36 V, respectively. Likely sharp redox peaks are observed at 5mV/s, which gradually smoothens at higher scan rates. This suggests for the pseudocapacitive property of the LiMn$_2$O$_4$ nanorods [55]. These redox peaks attribute to the redox activities of the electrolyte. As mentioned in the previous reports, the redox activity of electrolyte can be constituted to two different species, the one being from KOH and the other, K$_3$Fe(CN)$_6$. However in this electrolyte lithium ion movement is expected to be low but KOH electrolyte adds the redox kinetics in addition to the above lithiation/delithiation process (eq.1-2) through two concurrent activities [56, 57]:

$$Mn_2O_4 + K^+ + e^- \leftrightarrow (Mn_2O_4^- \ K^+)_{adsorption} \quad (4)$$

Equation (4) corresponds to the attachments of K$^+$ ions on the adsorbent surface (LiMn$_2$O$_4$ nanorods) without any chemical bonding relating non-faradaic activity. A simultaneous faradaic activity is also feasible involving intercalation-de-intercalation of electrons (e$^-$) and K$^+$ ions into the active sites of LiMn$_2$O$_4$ nanorods as presented:

$$Mn_2O_4 + K^+ + e^- \leftrightarrow Mn_2O_3.OK \quad (5)$$

The presence of redox additive K$_3$Fe(CN)$_6$ considerably adds up the redox activity by adding the redox coupled molecule [Fe(CN)$_6$]$^{3-}$/[Fe(CN)$_6$]$^{4-}$ [58]:

$$K_3Fe(CN)_6 + e^- \leftrightarrow K_4Fe(CN)_6 \quad (6)$$

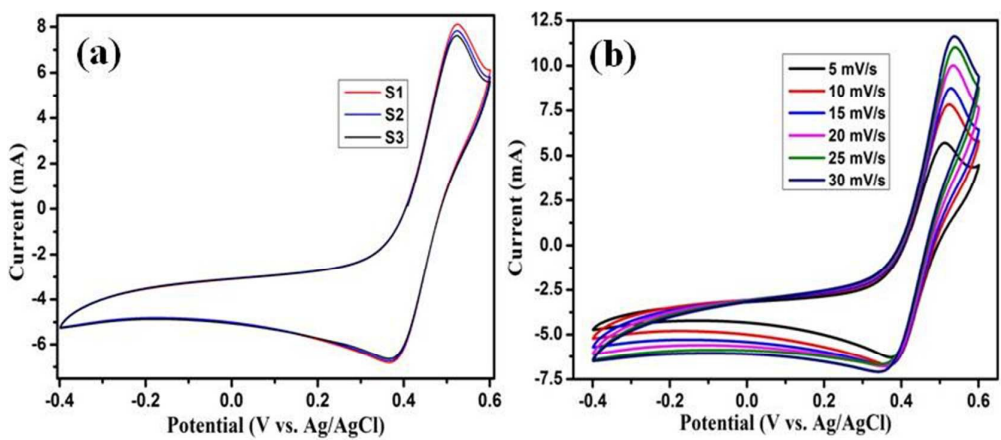

**Figure 9.** Cyclic voltammetry measurements for sample (a) S1, S2 and S3 at 15 mV/s and (b) S1 at different scan rates in KOH/K$_3$Fe(CN)$_6$ electrolyte.

The galvanostatic charge-discharge profiles for 1$^{st}$ cycle at promisingly high current densities of 30, 25, 20 and 15 A/g for sample S1 is displayed in fig. 10a. The specific capacities are calculated in similar way following eq. 3. The calculated capacities are 653.5, 651.5, 543.4, and 303.75 F/g at respective current densities of 15, 20, 25 and 30 A/g. In contrast, we observed inferior capacities in our previous reports of MnO$_2$ as 643.5, 270 and 185 F/g at current densities of 15, 20 and 25 A/g under same condition [55]. MnO$_2$ experiences comparable capacity at 15 A/g but considerable decrease in capacity at higher current density of 20 and 25 A/g with respect to LiMn$_2$O$_4$. This can be attributed to the lack of lithiation/de-lithiation processes in the former case. Likewise, the obtained capacities are more promising than the reports of 164 F/g at 16 A/g, 106 F/g at 0.5 A/g, 96 F/g at 20 A/g for MnO$_2$ nanorods [59], MnO$_2$/CNT composite [60] and MnO$_2$ thin film [61] respectively. Hence, a considerable enhancement in pseudo-capacitance is observed when the electrode is changed from α-MnO$_2$ nanorods to LiMn$_2$O$_4$ nanorods. This clearly supports that the as prepared lithiated manganese oxide (LiMn$_2$O$_4$) provides some extra redox activities in conjugation with those from the electrolyte. Retention capacity was tested for 4000 cycles at 30 A/g (fig. 10b). Stable cycling of first 20 cycles for sample S1 is highlighted in inset fig. 10b. During the initial cycling all the samples behave very closely. Negligible decay in capacity is observed for sample S1 compared to S2 and S3 upon prolong cycling. After 1000 cycles a noticeable decay is observed for sample S3 compared to S1 and S2. After 2000 cycles, the retention value for sample S2 also starts to show noticeable degradation in capacity against S1. Sample S1 performs the best with 93% of capacity retention after 4000 cycles followed by S2 and S3 which stands at 74% and 68.4%, respectively. This matches well with above analysis and gives weight-age to our assumption for the dependency of electrochemical storage property on morphology of electrode material. We predicted very high stability criteria for the electrode devoid of any rupture or pulverization in electrode material. However, a small decay in capacity could be attributed to the etching of electrode material due to improper coating. The study strongly recommends a considerable enhancement in charge storage property for LiMn$_2$O$_4$ in redox additive, KOH/K$_3$Fe(CN)$_6$ electrolyte compared to aqueous Li$_2$SO$_4$ electrolyte.

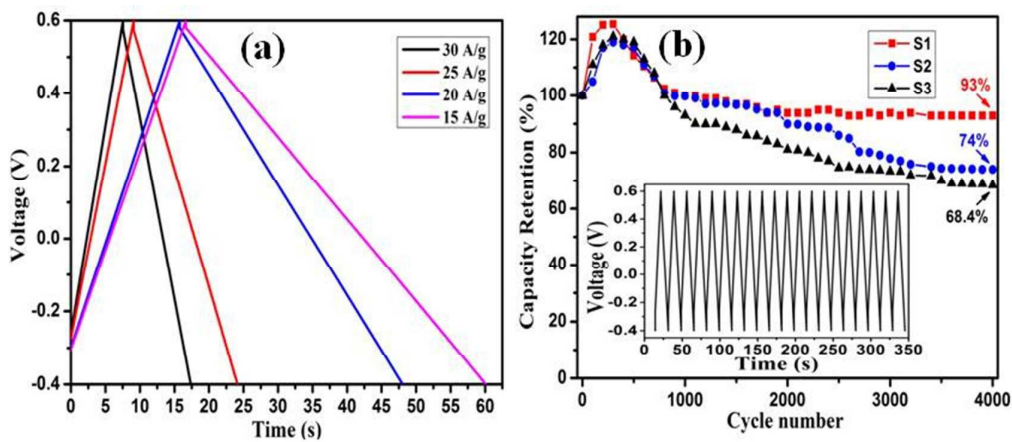

**Figure 10.** (a) Galvanostatic charge/discharge cycles at different current densities for S1 and (b) Capacity retention versus cycle number at 30 A/g for S1, S2 and S3 with inset (b) first 20 cycles of Galvanostatic charge/discharge for S1 in KOH/K$_3$Fe(CN)$_6$ electrolyte.

In supercapacitor more number of charge involvements is highly desirable for enhanced charge storage property. Compared to our previous report [55], the extra reversible charge activity relating lithiation/delithiation was absent and so a considerable less capacitance was encountered. The capacity increases during 100-500 cycles by maximum of 120% and then decreases to stable value exhibiting an excellent retention capacity of 93% (fig. 10b). This variation in retention capacity compared to the previous can be well attributed to the redox activity of the K$_3$Fe(CN)$_6$. During the initial charging process more number of Fe(II) converts to Fe(III) upon oxidation and during discharge process some of these Fe(III) converted back to Fe(II). The conversion of Fe(III) to Fe(II) is less compared to the formation of Fe(III) and so more number of charges (electrons) are available for redox activity, thereby, enhancing the charge storage capacity. However, after prolong cycling (500 cycles approx.), these charges are involved in reduction of Fe(III) to Fe(II) during discharge process and thus there number starts decreasing which in turn decreases the capacitive behaviour. Anyhow, during charging, again Fe(II) gets oxidized till a stage is reached, where the reversible conversion of Fe(II)↔Fe(III) attains equilibrium. This phenomenon helps to achieve a stable capacitive behaviour and thus a promising electrode material can be predicted in the form of LiMn$_2$O$_4$ nanorods under the redox added electrolyte. Acknowledging the complexity behind the pseudo-capacitive mechanism, a future study is required.

## 4. Conclusions

An effective synthesis procedure was successfully developed to achieve ultrafine LiMn$_2$O$_4$ nanorods with no physical imperfections using as grown $\beta$-MnO$_2$ nanorods. This included a novel hydrothermal method to homogenously mix the precursors which finally resulted in the lithiated nanorods of 10-80 nm diameters. BET method was effectively employed for surface analysis to account for ultrafine growth of LiMn$_2$O$_4$ nanorods with high surface area of 88.294 m$^2$/g. Ultrafine confined morphologies exhibited better electrochemical kinetics and reveals that charge storage capacity is dependent on morphology of the electrode material. Confined morphologies could facilitate the charge accumulation at electrode/electrolyte interfaces by providing short diffusion path. Furthermore, LiMn$_2$O$_4$ nanorods showed considerably better performance in KOH with K$_3$Fe(CN)$_6$ as electrolyte compared to aqueous Li$_2$SO$_4$ electrolyte with high pseudo-capacitance of 653.5 F/g at 15 A/g and stable capacity retention of 93% after 4000 cycles. It mainly stems from the enhanced redox activities of redox couple [Fe(CN)$_6$]$^{3-}$/[Fe(CN)$_6$]$^{4-}$ and K$^+$ ions of the electrolyte. This combination of electrode and electrolyte may prove to be beneficial for developing next generation of supercapacitors.

## Acknowledgements

This work was supported by DST Nano Mission, Govt. of India, via Project No. SR/NM/NS-1062/2012. The author T. Maiyalagan thanks the Department of Science and Technology-Science and Engineering Research Board [DST-SERB; No. ECR/2016/002025], India for financial support through Early Career Research Award.

## References

[1] Y.Z. Zhang, J. Zhao, J. Xia, L. Wang, W.Y. Lai, H. Pang, W. Huang, *Sci. Rep.*, 2015, **5**, 8536.
[2] X.-B. Cheng, H.-J. Peng, J.-Q. Huang, R. Zhang, C.-Z. Zhao, Q. Zhang, *ACS Nano*, 2015, **9**, 6373.
[3] J. Yan, Q. Wang, T. Wei, Z. Fan, *Adv. Energy Mater.*, 2014, **4**, 1300816.
[4] J. Yan, Q. Wang, C. Lin, T. Wei, Z. Fan, *Adv. Energy Mater.*, 2014, **4**, 1400500.
[5] X. Tian, X. Sun, Z. Jiang, Z.-Jie Jiang, X. Hao, D. Shao, T. Maiyalagan, *ACS Appl. Energy Mater.*, 2018, **1**, 143.
[6] H. Tanaya Das, K. Mahendraprabhu, T. Maiyalagan, P. Elumalai, *Sci. Rep.*, 2017, **76**, 15342 .
[7] X. Wang, Y. Hou, Y. Zhu, Y. Wu, R. Holze, *Sci. Rep.*, 2013, **3**, 1401.
[8] W. Jiang, D. Yu, Q. Zhang, K. Goh, L. Wei, Y. Yong, R. Jiang, J. Wei, Y. Chen, *Adv. Funct. Mater.*, 2015, **25**, 1063.
[9] D. Cai, H. Huang, D. Wang, B. Liu, L. Wang, Y. Liu, Q. Li, T. Wang, *ACS Appl. Mater. Interfaces*, 2014, **6**, 15905.
[10] H.L. Wang, Y.Y. Liang, T. Mirfakhrai, Z. Chen, H.S. Casalongue, H.J. Dai, *Nano Res.*, 2011, **4**, 729.

[11] Md. Uddin, H. Tanaya Das, T. Maiyalagan, P. Elumalai *Appl. Surf. Sci.,* 2018, **449**, 445-453.
[12] H. Chen, S. Zhou, L. Wu, *ACS Appl. Mater. Interfaces*, 2014, **6**, 8621.
[13] K. Leng, F. Zhang, L. Zhang, T.F. Zhang, Y.P. Wu, Y.H. Lu, Y. Huang, Y.S. Chen, *Nano Res.*, 2013, **6**, 581.
[14] M.-Q. Zhao, Q. Zhang, J.-Q. Huang, G.-L. Tian, T.-C. Chen, W.-Z. Qian, F. Wei, *Carbon*, 2013, **54**, 403.
[15] D. Zhou, H. Lin, F. Zhang, H. Niu, L. Cui, Q. Wang, F. Qu, *Electrochim. Acta*, 2015 **161,** 427.
[16] X. Du, C. Wang, M. Chen, Y. Jiao, J. Wang, *J. Phys. Chem. C*, 2009, **113**, 2643-2646.
[17] W. Tang, L.L. Liu, S. Tian, L. Li, Y.B. Yue, Y.P. Wu, K. Zhu, *ChemComm*, 2011, **47**, 10058-10060.
[18] X. Yanga, F. Qua, H. Niua, Q. Wanga, J. Yanb, Z. Fanb, *Electrochim. Acta*, 2015, **180**, 287–294.
[19] H.H. Xu, X.L. Hu, H.L. Yang, Y.M. Sun, C.C. Hu, Y.H. Huang, *Adv. Energy Mater.*, 2015, **5**, 1401882.
[20] F. Gao, J.Y. Qu, Z.B. Zhao, Q. Zhou, B.B. Li, J.S. Qiu, *Carbon*, 2014, **80,** 640.
[21] S.S. Li, Y.H. Luo, W. Lv, W.J. Yu, S.D. Wu, P.X. Hou, Q.H. Yang, Q.B. Meng, C. Liu, H. M. Cheng, *Adv. Energy Mater.*, 2011, **1,** 486.
[22] R.R. Salunkhe, J. Lin, V. Malgras, S.X. Dou, J.H. Kim, Y. Yamauchi, *Nano Energy*, 2015, **11**, 211.
[23] P. Tang, L. Han, L. Zhang, *ACS Appl. Mater. Interfaces*, 2014, **6**, 10506.
[24] G. Sun, X. Zhang, R. Lin, J. Yang, H. Zhang, P. Chen, *Angew. Chem. Int. Ed.*, 2015, **54**, 4651.
[25] T. Zhai, S.L. Xie, M.H. Yu, P.P. Fang, C.L. Liang, X.H. Lu, Y.X. Tong, *Nano Energy*, 2014, **8**, 255.
[26] W. Wang, W.Y. Liu, Y.X. Zeng, Y. Han, M.H. Yu, X.H. Lu, Y.X. Tong, *Adv. Mater.* 2015, **27**, 3572.
[27] B.H. Zhang, Y. Liu, Z. Chang, Y.Q. Yang, Z.B. Wen, Y.P. Wu, R. Holze, *J. Power Sources*, 2014, **253**, 98.
[28] J. Liu, Y. Shen, L. Chen, Y. Wang, Y. Xia, *Electrochim. Acta*, 2015, **156**, 38.
[29] X. Jin, Q. Xu, X. Yuan, L. Zhou, Y. Xia, *Electrochim. Acta*, 2013, **114**, 605.
[30] S. Lee, G. Yoon, M. Jeong, M.J. Lee, K. Kang, J. Cho, *Angew. Chem. Int. Ed.*, 2015, **54**, 1153.
[31] F. Han, L.J. Ma, Q. Sun, C. Lei, A.H. Lu, *Nano Res.*, 2014, **7**, 1706.
[32] L. Mai, H. Li, Y. Zhao, L. Xu, X. Xu, Y. Luo, Z. Zhang, W. Ke, C. Niu, Q. Zhang, *Sci. Rep.*, 2013, **3**, 1718.
[33] G.J. Wang, Q.T. Qu, B. Wang, Y. Shi, S. Tian, Y.P. Wu, *ChemPhysChem*, 2008, **9**, 2299.
[34] J. Lu, C. Zhou, Z. Liu, K. S. Lee, L. Lu, *Elctrochim. Acta*, 2016, **212**, 553-560.
[35] Y. Hou, Y. Cheng, T. Hobson, J. Liu, *Nano Lett.*, 2010, **10**, 2727-2733.
[36] W. Tang, S. Tian, L.L. Liu, L. Li, H.P. Zhang, Y.B. Yue, Y. Bai, Y.P. Wu, K. Zhu, *Electrochem. Commun.*, 2011, **13**, 205.
[37] M. Zhao, X. Song, F. Wang, W. Dai, X. Lu, *Electrochim. Acta*, 2011, **56**, 5673-5678.
[38] Q.T. Qu, L.J. Fu, X.Y. Zhan, D. Samuelis, J. Maier, L. Li, S. Tian, Z.H. Li, Y.P. Wu, *Energy Environ. Sci.*, 2011, **4**, 3985-3990.
[39] W. Tang, Y. Hou, F. Wang, L. Liu, Y. Wu, K. Zhu, *Nano Lett.*, 2013, **13**, 2036-2040.
[40] M. A. Kebede, K. I. Ozoemena, *Mater. Res. Express*, 2017, **4**, 025030.
[41] M. Fehse, R. Trocoli, E. Ventosa, E. Hernandez, A. Sepulveda, A. Morata, A. Tarancon, *ACS Appl. Mater. Interfaces*, 2017, **9 (6)**, 5295–5301.

[42] B. K. Lesel, J. S. Ko, B. Dunn, S. H. Tolbert, *ACS Nano*, 2016, **10**, 7572−7581.
[43] F.X. Wang, S.Y. Xiao, Y.S. Zhu, Z. Chang, C.L. Hu, Y.P. Wu, R. Holze, *Journal of Power Sources*, 2014, **246**, 19-23.
[44] N. Kumar, P. Dineshkumar, R. Rameshbabu, A. Sen, *RSC Adv.*, 2016, **6**, 7448–7454.
[45] N. Kumar, P. Dineshkumar, R. Rameshbabu, A. Sen, *Mater. Lett.*, 2015, **158**, 309-312.
[46] N. Kumar, A. Sen, K. Rajendran, R. Rameshbabu, J. Ragupathi, H. Therese, T. Maiyalagan, *RSC Adv.*, 2017, **7**, 25041-25053.
[47] N. Kumar, S. Bhaumik, A. Sen, A. P. Shukla, S. D. Pathak, *RSC Adv.*, **7**, 34138-34148.
[48] G. Kumar, Z. Awan, K. S. Nahm, J. S. Xavier, *Biosens. Bioelectron.*, 2014, **53**, 528–534 (2014).
[49] M. Hu, K.S. Hui, K.N. Hui, *Chem. Eng. J.*, 2014, **254**, 237–244.
[50] C. V. Ramana, M. Massot, C. M. Julien, *Surf. Interface Anal.*, 2005, **37**, 412–416.
[51] Julien C, Massot M, Rangan S, Lemal M, Guyomard D., *J. Raman Spectrosc.*, 2002, **33**, 223.
[52] G. Cohn, *Chem. Rev.*, 1948, **42** (3), 527–579.
[53] M. Kruk, M. Jaroniec, *Chem. Mater.*, 2001, **13**, 3169-3183.
[54] Q. Liu, J. Yang, R. Wang, H. Wang, S. Ji, *RSC Adv.*, 2017, **7**, 33635.
[55] N. Kumar, K. Prasad, A. Sen, T. Maiyalagan, *Appl. Surf. Sci.,* 2018, **449**, 492-499.
[56] M. Toupin, T. Brousse, D. Belanger, *Chem. Mater.*, 2004, **16**, 3184-3190.
[57] S. C. Pang, M. A. Anderson, T. W. Chapman, *Journal of The Electrochemical Society*, 2000, **147 (2)**, 444-450.
[58] K. Chen, S. Song, D. Xue, *RSC Adv.*, 2014, **4**, 23338-23343.
[59] X. Su, X. Yang, L. Yu, G. Cheng, H. Zhang, *CrystEngComm*, 2015, **17**, 5970-5977.
[60] D.Z.W. Tan, H. Cheng, S.T. Nguyen, H.M. Duong, *Mater. Technol.*, 2014, **29**, A107–A113.
[61] Y. Zhao, P. Jiang, S. S. Xie, *J. Power Sources*, 2013, **239**, 393.